\documentclass[aps,prb,showpacs,twocolumn]{revtex4-1}

\usepackage{natbib}
\usepackage{graphicx}
\usepackage[ansinew]{inputenc}

\begin{document}

\title{Kondo effect of a Co atom on Cu(111) in contact with an Fe tip}
\author{N. Néel} 
\author{J. Kröger}
\email{joerg.kroeger@tu-ilmenau.de} 
\altaffiliation[Present address: ]{Institut für Physik, Technische Universität 
Ilmenau, D-98693 Ilmenau, Germany}
\author{R. Berndt}
\affiliation{Institut für Experimentelle und Angewandte Physik, 
Christian-Albrechts-Universität zu Kiel, D-24098 Kiel, Germany}

\begin{abstract}
Single Co atoms, which exhibit a Kondo effect on Cu(111), are contacted with 
Cu and Fe tips in a low-temperature scanning tunneling microscope. With Fe tips,
the Kondo effect persists with the Abrikosov-Suhl resonance significantly 
broadened. In contrast, for Cu-covered W tips, the resonance width remains 
almost constant throughout the tunneling and contact ranges. The distinct 
changes of the line width are interpreted in terms of modifications of the Co 
$d$ state occupation owing to hybridization with the tip apex atoms.
\end{abstract}


\maketitle

The Kondo effect, one of the key correlation effects in condensed matter 
physics, results from scattering of conduction electrons from a 
localized half-filled $d$ or $f$ electron level. Below a characteristic Kondo 
temperature ($T_{\text{K}}$) the excitation spectrum of the local impurity 
exhibits an Abrikosov-Suhl resonance at the Fermi level with a width 
$\text{k}_{\text{B}}T_{\text{K}}$ ($\text{k}_{\text{B}}$: Boltzmann's 
constant). Its lineshape is susceptible to changes of 
the local environment of the impurity and to magnetic fields. In external 
magnetic fields, Zeeman splitting of the electron level involved in the Kondo 
effect leads to splitting of the resonance. 
\cite{dgo_98,scr_98, ahe_04,aot_08} Addtional effects occur in magnetic 
nanostructures. Interactions between two Kondo impurities mediated by 
itinerant electrons may result in a complete or partial suppression or a 
splitting of the resonance, depending on the relative magnitudes of the Kondo 
energy scales of the individual impurities and the interaction energy. 
\cite{wch_99,*tja_01,ncr_04,jch_04,psi_05,mva_05,hhe_06,pwa_07,aot_09} A 
splitting of the resonance observed from impurities at ferromagnetic islands 
has been attributed to exchange interaction. \cite{ska_10} Surprisingly, a 
Kondo effect has even been reported from atomic point contacts between 
ferromagnetic electrodes, which implies a significantly modified  magnetism. 
\cite{mca_09} Recent theoretical work showed that ferromagentic coupling
between nearly ferromagnetic Pd or Pt electrodes and a magnetic impurity 
may lead to an Abrikosov-Suhl resonance or a pseudogap at the Fermi level
depending on the coupling strength. \cite{pge_09} The Abrikosov-Suhl
resonance may also be tuned by controlling the hybridization of a magnetic
impurity with nonmagnetic neighbors. \cite{nne_07,lvi_08,nne_08}

Here, we use a scanning tunneling microscope (STM) to directly compare the 
evolution of the Kondo effect of a single Co atom adsorbed on Cu(111) upon 
approaching a ferromagnetic Fe and a nonmagnetic Cu-covered W tip until contact 
is reached. We find that hybridization effects have to be considered for
ferromagnetic tips to adequately describe the line shape of the Abrikosov-Suhl
resonance. Approaching the adsorbed atom (adatom) with Fe tips shows that the 
Abrikosov-Suhl resonance persists at contact. No splitting is resolved, but the 
resonance is significantly broadened. In contrast, Cu-covered W tips leave the 
resonance almost unaffected throughout the tunneling and contact ranges. As the 
resonance line shape is linked to the occupation of the Co $d$ states, these 
results may be interpreted in terms of different hybridizations between the Co 
atom and the tip materials without involving magnetic interactions. So far, 
little is known from experiments about the electronic states of an atom or a 
molecule in contact with two electrodes. \cite{nne08b,nne08c} This modest data 
base on basic electronic (and structural) properties is in some contrast to a 
wealth of theoretical results on, {\it e.\ g.}, energy-resolved conductances. 
The detour via the Kondo effect may therefore turn out to be useful in honing 
electronic structure calculations, which underly the transport results.

The experiments were performed with a home-made STM operated at $7\,\text{K}$ 
and in ultrahigh vacuum ($10^{-9}\,\text{Pa}$). Tips were fabricated from pure
polycrystalline W and Fe wires by chemical etching in solutions of NaOH and
HCl, respectively. Sample surfaces and Fe and W tips were cleaned by Ar$^+$ 
bombardment and annealing. Prior to mounting to the cold STM Fe tips were 
placed close to a CoSm permanent magnet to induce a magnetization along the 
long tip axis. Co atoms were deposited onto the Cu(111) surface at 
$\approx 10\,\text{K}$ using an electron beam evaporator and an evaporant of 
$99.99\,\%$ purity. Spectroscopy of the differential conductance 
($\text{d}I/\text{d}V$) was performed by a lock-in technique with a modulation 
of $1\,\text{mV}_{\text{rms}}$ and $8\,\text{kHz}$ added to the sample voltage. 
Conductance versus tip displacement curves were aquired by moving the tip 
towards the adatom at a velocity of $50\,\text{\AA}\text{s}^{-1}$ and 
simultaneously recording the current. The cleanliness of the tips was checked 
by $\text{d}I/\text{d}V$ spectra, which exclusively showed the spectroscopic 
signature of the Cu(111) surface state on pristine Cu(111). Particular care 
was taken to avoid coating of the Fe tip apex with substrate material.
\begin{figure}
  \includegraphics[width=75mm]{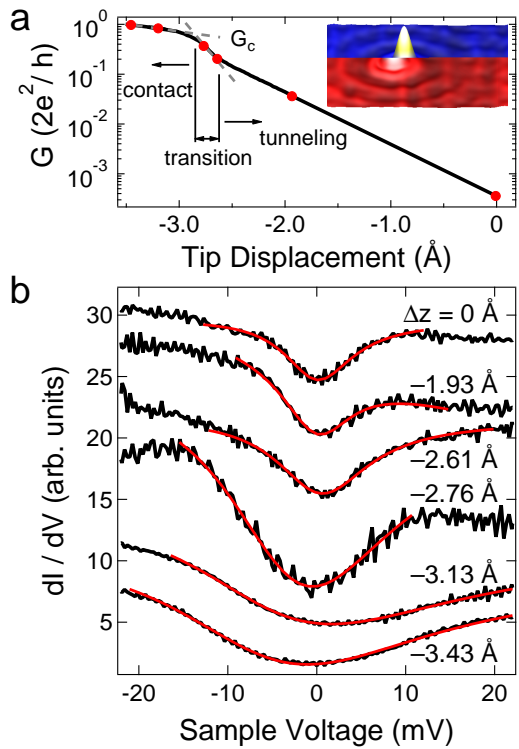}
  \caption[fig1]{(Color online) (a) Conductance of a Co adatom on Cu(111) 
  approached by a Fe tip. Tunneling, transition and contact ranges are 
  indicated. The conductance trace between $0$ and $\approx -2\,\text{\AA}$ is
  an exponential extrapolation of tunneling conductance in the displacement
  range from $\approx -2$ to $-2.5\,\text{\AA}$. The contact conductance, 
  $G_{\text{c}}$, is defined by the intersection of linear fits (dashed lines) 
  to conductance data in the contact and transition ranges. Zero displacement 
  is defined by feedback loop parameters of $33\,\text{mV}$ and $1\,\text{nA}$. 
  Inset: Pseudo-three-dimensional presentation of a constant-current STM image 
  of a single Co adatom on Cu (111) ($8\,\text{\AA}\times 8\,\text{\AA}$, 
  $33\,\text{mV}$, $1\,\text{nA}$). (b) Spectra of $\text{d}I/\text{d}V$ 
  acquired at junction conductances indicated by dots on the conductance trace 
  in (a). Solid lines are fits of Fano line shapes to experimental data.}
  \label{fig1}
\end{figure}
  
Figure \ref{fig1}(a) shows the evolution of the conductance of a single-Co 
junction upon approaching (from right to left) a Fe tip to the individual Co 
adatom on Cu(111). The tunneling range is characterized by an exponential
variation of the conductance with the displacement, $\Delta z$
($0\,\text{\AA}>\Delta z>-2.65\,\text{\AA}$). The transition from tunneling to
contact takes place within $\approx 0.35\,\text{\AA}$
($-2.65\,\text{\AA}>\Delta z>-3\,\text{\AA}$). For $\Delta z<-3\,\text{\AA}$ 
the contact range is reached and the conductance rises slowly with further tip 
displacement. A contact conductance, $G_{\text{c}}$, is defined as the 
intersection of linear fits to conductance data in the contact and transition 
ranges. \cite{jkr_08} The single-Co contact exhibits a conductance of 
$G_{\text{c}}\approx(0.70\pm 0.05)\,\text{G}_0$ 
($\text{G}_0=2\text{e}^2/\text{h}$, e: electron charge, h: Planck's constant).
This value is lower than the quantum of conductance, $\text{G}_0$, which has 
been obtained for Co adatoms contacted with nonmagnetic electrodes. 
\cite{nne_07,lvi_08} However, it is in agreement with the contact conductance 
measured from a single Co adatom contacted by a ferromagnetic Ni tip and a 
Cu(111) surface. \cite{nne_09}

In a next step, the tip position was frozen at characteristic displacements
[indicated by dots on the conductance trace in Fig.\,\ref{fig1}(a)] and spectra 
of $\text{d}I/\text{d}V$ were acquired. The resulting data are presented in
Fig.\,\ref{fig1}(b). The three top spectra were recorded in the tunneling
range, the fourth spectrum in the transition between tunneling and contact,
and the two lower spectra were acquired in the contact range. Prior to and 
after contact the Co adatom was imaged and $\text{d}I/\text{d}V$ spectroscopy
was performed to detect possible modifications of structural or electronic 
properties of the junction. The spectroscopic signature of the Kondo effect is 
present in all $\text{d}I/\text{d}V$ spectra and appears with a Fano line shape
around zero voltage. \cite{jli_98,vma_98} While the tunneling spectra exhibit 
similar line shapes, the Abrikosov-Suhl resonance starts to broaden in the 
transition range and appears even wider in contact. The observation of a 
broadened Abrikosov-Suhl resonance of the Co adatom in contact with a 
ferromagnetic Fe tip is surprising in the light of previous work that reported 
the disappearance or splitting of the resonance when the Kondo impurity was 
close to magnetic atoms \cite{wch_99,*tja_01,pwa_07,aot_09} or nanostructures. 
\cite{ska_10} The only work showing the presence of the (unsplit) 
Abrikosov-Suhl resonance in ferromagnetic point contacts has been reported by
Reyes {\it et al.}\,\cite{mca_09} From a statistical analysis of 
$\text{d}I/\text{d}V$ spectra the authors of Ref.\,\onlinecite{mca_09} inferred 
that the Kondo effect is present in ferromagnetic Fe, Co and Ni contacts. They 
suggested that reduced symmetry and the decreased coordination of the atom in 
the contact may favor the observation of the Kondo effect. 

\begin{figure}
  \includegraphics[width=85mm]{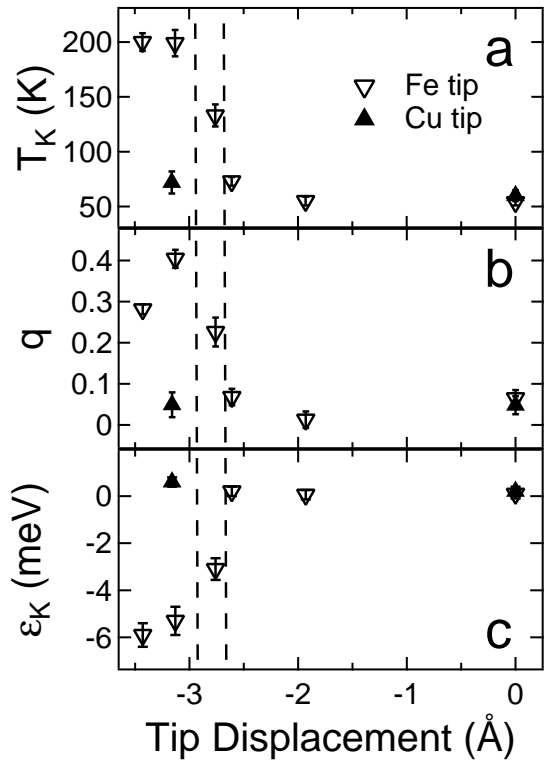}
  \caption[fig2]{Kondo temperature $T_{\text{K}}$, asymmetry factor $q$ and
  resonance energy $\epsilon_{\text{K}}$ for a Co adatom on Cu(111) contacted
  with a Fe (open symbols) and a Cu-covered W (filled triangles) tip as a 
  function of the tip displacement. Uncertainty margins result from fitting a 
  variety of $\text{d}I/\text{d}V$ spectra. Dashed lines separate tunneling, 
  transition and contact ranges as introduced in Fig.\,\ref{fig1}(a).}
  \label{fig2}
\end{figure}

To analyse the modifications of the Abrikosov-Suhl resonance at contact, a Fano 
line shape may be used. \cite{nne_07,dja_09} The resulting parameters, 
{\it i.\,e.}, Kondo temperature $T_{\text{K}}$, resonance energy 
$\epsilon_{\text{K}}$ and asymmetry parameter $q$, are presented in 
Fig.\,\ref{fig2} for spectra acquired with a Fe (open symbols) and a Cu-covered 
W (filled triangles) tip. In the tunneling range ($\Delta z>-2.65\,\text{\AA}$), 
these parameters are almost equal and constant for both tip materials, which 
reflects that the interaction between the tip and the sample are negligible in 
this conductance range. Starting from the transition range, however, Fe and 
Cu-covered W tips lead to strikingly different results. For Fe tips, all 
parameters start to deviate from their tunneling range values. $T_{\text{K}}$
increases from $\approx 60\,\text{K}$ to $\approx 130\,\text{K}$, $q$ increases 
from $\approx 0.05$ to $\approx 0.25$, and the resonance energy drops from 
$\approx 0$ to $\approx -3\,\text{meV}$. At contact, this trend is continued 
and $T_{\text{K}}$ reaches $\approx 200\,\text{K}$, $q\approx 0.3$ - $0.4$ and 
$\epsilon_{\text{K}}\approx -6\,\text{meV}$. For Cu-coated W tips all 
parameters are essentially constant throughout the tunneling, transition and 
contact ranges in agreement with previous results. \cite{lvi_08} Obviously, the 
chemical identity of the tip apex atom determines the degree of hybridization.
The splitting of the resonance due to the magnetic dipole field of the Fe tip 
at the adatom site is too low to explain the observed broadening. Estimating 
the dipole field at contact as $H\approx 1\,\text{T}$, \cite{aku_02} the 
splitting is of the order of $2\text{g}\mu_{\text{B}}H\approx 0.2\,\text{meV}$ 
(g: Land\'{e} factor, $\mu_{\text{B}}$: Bohr's magneton), \cite{tco_00} which 
is more than one order of magnitude lower than 
$\text{k}_{\text{B}}T_{\text{K}}$ with $T_{\text{K}}\approx 60\,\text{K}$.

The asymmetry parameter $q$ describes the coupling of the probe to the discrete 
Co $d$ state and the continuum of $sp$ conduction electrons. 
\cite{jli_98,vma_98} The considerable increase of $q$ upon hybridization of the 
Co adatom with the Fe tip suggests that additional tip states couple to the Co 
$d$ levels. Indeed, ferromagnetic Fe exhibits $d$ bands at the Fermi level, 
\cite{atu_84,*tna_86} which may hybridize with the adatom $d$ states at 
contact. The $d$ bands of Cu, however, are well below the Fermi energy 
\cite{wkr_69,*nsm_69} and thus no additional hybridization is expected in 
agreement with an almost constant asymmetry parameter. In the tunneling range,
owing to the stronger spatial decay of $d$ states compared to $s$ states, 
\cite{mha_08} only $s$ states participate in the hybridization and thus lead to 
a similar $q$ parameter for Fe and Cu-covered W tips. 

\begin{figure}
  \includegraphics[width=85mm]{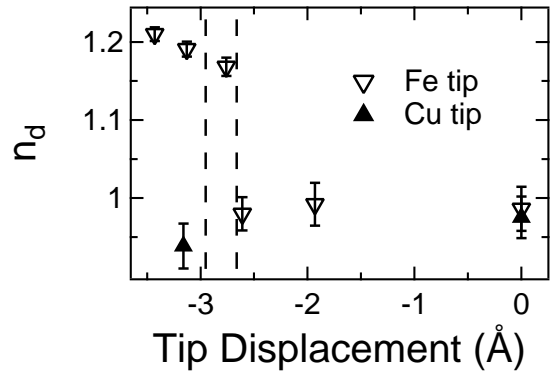}
  \caption[fig3]{Occupation number of Co $d$ states, $n_d$, as a function of 
  the tip displacement. Open (filled) triangles present data for a Co adatom on
  Cu(111) contacted by a Fe (Cu-covered W) tip. $n_d$ has been calculated using
  Eq.\,(\ref{nd}) with fit parameters $\epsilon_{\text{K}}$ and $T_{\text{K}}$
  (Fig.\,\ref{fig2}). Dashed lines separate tunneling, transition and 
  contact ranges as introduced in Fig.\,\ref{fig1}(a).}
  \label{fig3}
\end{figure}

The different hybridization may be described by the Co $d$ state occupation 
number, $n_d$, which is related to the resonance energy and the Kondo 
temperature via \cite{ahe_93}
\begin{equation}
  n_d = 1 - \frac{2}{\pi}\,\tan^{-1}
  \left(
  \frac{\epsilon_{\text{K}}}{\text{k}_{\text{B}}\text{T}_{\text{K}}}
  \right)
  \label{nd}
\end{equation}
and thus may be estimated from the extracted fit parameters. In the following
we assume that only a single Co $d$ level participates in the hybridization.
This assumption appears to be reasonable in the light of a recent theoretical
study of a Co adatom on Au(111). \cite{ouj_00} Owing to $sp$-$d$ hybridzation
four Co $d$ orbitals are completely occupied while the orbital which is 
responsible for the Kondo effect is partly filled with $0.8$ electrons.
\cite{ouj_00} Empty, half-filled, and filled $d$ levels correspond to $n_d=0$, 
$1$, and $2$, respectively. Figure \ref{fig3} shows $n_d$ according to 
Eq.\,(\ref{nd}) as a function of the tip displacement for Fe (open symbols) and 
Cu-covered W (filled triangles) tips. Clearly, $n_d$ changes upon hybridization of 
the Co adatom with the Fe tip apex atom from $\approx 0.98$ (average value in 
the tunneling range) to $\approx 1.2$ (contact) while it stays almost constant 
for the Cu-covered W tip ($\approx 0.98$). A value of $\approx 0.98$ is in good 
agreement with the occupation number obtained for Co adatoms on Cu(111) in the 
tunneling range, while $n_d\approx 1.2$ comes close to the value of a Co adatom 
on Cu(100). \cite{pwa_04}

The influence of $n_d$ on $T_{\text{K}}$ may be evaluated according to 
\cite{ouj_00}
\begin{equation}
  \text{k}_{\text{B}}T_{\text{K}}\approx 
  \sqrt{\frac{\Delta U}{2}}
  \exp\left(
  -\frac{\pi}{2}\frac{U}{\Delta}
  \left|
  \frac{3}{2}-n_{d}
  \right|
  \left|
  \frac{1}{2}-n_{d}
  \right|
  \right),
  \label{kondo}
\end{equation}
where $U$ is the on-site Coulomb repulsion and $\Delta$ the half width of the 
hybridized $d$ level. Using Eq.\,(\ref{kondo}) with $U=2.4\,\text{eV}$ and 
$\Delta =0.2\,\text{eV}$ as calculated for Co adatoms on Cu(100) in the 
tunneling range, \cite{nne_07} a Kondo temperature of 
$T_{\text{K}}\approx 50\,\text{K}$ is obtained for $n_d=0.98$. At contact, 
the on-site Coulomb repulsion is reduced to $U=1.9\,\text{eV}$ as previously
reported in Ref.\,\onlinecite{nne_07}. Together with the increased occupation
number $n_d=1.2$ at contact (Fig.\,\ref{fig3}) a Kondo temperature of 
$T_{\text{K}}\approx 200\,\text{K}$ is obtained. These values are in agreement 
with the measured Kondo temperatures in the tunneling and contact ranges 
[Fig.\,\ref{fig2}(a)]. Consequently, the experimentally observed variations of 
$T_{\text{K}}$, $\epsilon_{\text{K}}$ and $q$ are compatible with variations of 
the Co $d$ level occupation number upon hybridization with the tip and do not 
rely on a ferromagnetic exchange interaction between the tip apex and the 
adsorbed atom. The occupation number may likewise be altered by a modification 
of the hybridization with the substrate. At contact, owing to adhesive forces 
between the tip and the adatom, the adatom is lifted from the surface, 
\cite{rzh_06,lvi_08} which in turn affects the $d$ level occupation.

In conclusion, the Kondo effect of a single magnetic impurity has been used to 
monitor the hybridization of the impurity with Cu and Fe tips. The degree of 
hybridization has been inferred from the impurity $d$ level occupation number, 
which in turn has been extracted from the width of the Abrikosov-Suhl 
resonance.

Financial support by the Deutsche Forschungsgemeinschaft through SFB 668 is 
ackowledged.

\end{document}